\documentstyle[prd,aps,preprint,tighten,floats,eqsecnum,epsf]{revtex} 

\input{epsf}

\begin{document}

\reversemarginpar

\title{Fluctuations in the current and energy densities\\ 
around a magnetic flux carrying cosmic string}
\author{L.~Sriramkumar~\thanks{E-mail:~slakshm@phys.ualberta.ca}}
\address{Theoretical Physics Institute, Department of Physics\\
University of Alberta, Edmonton, Alberta~T6G~2J1, Canada}

\maketitle

\begin{abstract}
We calculate the fluctuations in the current and energy densities 
for the case of a quantized, minimally coupled, massless, complex
scalar field around a straight and infinitesimally thin cosmic string 
carrying magnetic flux. 
At zero temperature, we evaluate the fluctuations in the current
and energy densities for arbitrary flux and deficit angle.
At a finite temperature, we evaluate the fluctuations in the
energy density for the special case wherein the flux is absent 
and the deficit angle equals~$\pi$. 
We find that, quite generically, the dimensionless ratio of the 
variance to the mean-squared values of the current and energy
densities are of order unity which suggests that the fluctuations 
around cosmic strings can be considered to be large.
\end{abstract}
\newpage

\section{Introduction}\label{sec:intro}

During the last two decades or so, there has been considerable 
interest in literature in delineating the domain of validity of 
semiclassical gravity wherein the backreaction of a quantum field 
on a classical gravitational background is assumed to be given 
by the expectation value of the stress-energy tensor of the quantum 
field (see Refs.~\cite{ford82,paddytp90}; for a detailed discussion 
and further references, see, for e.g., Ref.~\cite{paddytp89}).
With this motivation in mind, the fluctuations in the stress-energy 
densities of quantum fields have been evaluated in a variety of 
situations in flat spacetime~\cite{kf93,ph97,wf99} and, to a 
relatively limited extent, in certain curved spacetimes as 
well~\cite{ph97,wf99,ph00b}.
The different possible ways of regularizing the four-point 
functions that one encounters when evaluating the fluctuations 
in the stress-energy density have been 
considered~\cite{kf93,ph97,wf99,ba91a,ba91b,eb92,hp00,ph00a} 
and the possible implications of these results on the validity 
of semiclassical gravity have also been discussed in some 
detail (see Refs.~\cite{kf93,ph97,wf99}; for a critical outlook, 
see Refs.~\cite{hp00,ph00a}).

Apart from the cases that have already been considered 
in literature, the non-trivial spacetime around a cosmic 
string~\cite{vi81,vi85,go85,hi85} provides another 
interesting (and easily tractable) situation to study 
the fluctuations in the stress-energy densities of quantum
fields.
The mean values of the stress-energy tensor for different
quantum fields around a cosmic string have been evaluated at both 
zero~\cite{do77,hk86,li87,fs87,do87a,do87b,do90,sm90,gl94,mo95} 
and at a finite temperature~\cite{ds88,li92,ro93,fpz95}. 
In fact, situations wherein the cosmic string carries 
a non-zero magnetic flux have been considered as 
well~\cite{fs87,do90,sm90,gl94,gu95}. 
It is well-known that a Aharanov-Bohm solenoid induces a 
non-zero current around it~\cite{se85,go90,bo91,pa91,sb98,si99}. 
Therefore, a flux carrying cosmic string, in addition to inducing 
a non-zero stress-energy density, will also induce a non-zero 
current.
Our aim in this paper is to evaluate the fluctuations in 
the current and energy densities of a quantum field around 
a flux carrying cosmic string.
The quantum field we shall consider is a minimally coupled, 
massless, complex scalar field and we shall evaluate the 
fluctuations in two situations.
At zero temperature, we shall evaluate the fluctuations in the 
current and energy densities for arbitrary flux and deficit angle.
At a finite temperature, we shall evaluate the fluctuations in the 
energy density for the special case wherein the flux is absent and 
the deficit angle equals~$\pi$. 

This paper is organized as follows.
In the following section, we shall briefly re-derive the 
two-point function for the quantized, massless, complex 
scalar field around the flux carrying cosmic string at 
zero temperature.
We shall also obtain the finite temperature two-point 
function for the special case wherein the flux is absent 
and the deficit angle of the cosmic string equals~$\pi$. 
Using these two-point functions, we shall obtain the mean 
values of the four-current and stress-energy densities.
In Sec.~\ref{sec:flctns}, using Wick's theorem, we shall 
express the four-point functions of the quantum field in 
terms of the two-point functions and evaluate the 
fluctuations in the current and energy densities.
We shall also discuss the regularization procedure we 
have adopted in order to obtain finite expressions for 
the fluctuations.   
In Sec.~\ref{sec:mag}, we shall evaluate the relative magnitude 
of the variance (i.e. the mean-squared deviations) with respect 
to the mean-squared values of the current and energy densities
and briefly comment on the results we have obtained.
We shall close with Sec.~\ref{sec:dscssn} wherein we motivate 
the need to evaluate the mean values and the fluctuations around
cosmic strings using smeared fields or at separated points rather 
than evaluating them at the same spacetime point as we have done 
in this paper.

Before we proceed further, the following comments on our 
conventions and notations are in order. 
We shall set $\hbar=G=c=1$ and we shall denote the single 
unit of electric charge as~$e$.
We shall work in $(3+1)$ dimensions with a Lorentzian metric 
signature of $(+,-,-,-)$. 
For the sake of convenience and clarity in notation, we shall 
denote the set of four coordinates $x^{\mu}$ as ${\tilde x}$ 
and we shall write the derivatives $(\partial/\partial x)$ simply 
as $\partial_x$. 
Finally, we shall denote complex conjugation and Hermitian 
conjugation by an asterisk and a dagger, respectively.

\section{Two-point functions and mean values}\label{sec:tpfmv}

The spacetime around a straight and infinitesimally thin cosmic 
string that is oriented along the $z$-direction is described by 
the line element~\cite{vi81,vi85,go85,hi85}
\begin{equation}
ds^2 =dt^2-d\rho^2-\alpha^2 \rho^2 d\phi^2 -dz^2,\label{eqn:csmtrc}
\end{equation}
where $\rho > 0$, $0 \le \phi < (2\pi)$ and $\alpha=(1-4{\bar \mu})$, 
${\bar \mu}$ being the mass per unit length of the string. 
The line-element~(\ref{eqn:csmtrc}), though locally flat, is 
not so globally. 
The presence of the string leads to conical singularity and, 
as a result, the spacetime exhibits an azimuthal deficit angle 
of~$\left[2\pi(1-\alpha)\right]$.
Clearly, $\alpha=1$ corresponds to Minkowski spacetime. 

Let us now assume that the infinitesimally thin cosmic string 
carries an internal magnetic flux~${\bar \gamma}$.
Such a magnetic flux can be described by the vector 
potential~\cite{se85,go90,bo91,pa91,sb98,si99}
\begin{equation}
A_{\mu}=B\, \left(\partial_{\mu}\phi\right),\label{eqn:vp}
\end{equation}
where $B$ is a constant. 
This vector potential is singular along the string and the 
constant~$B$ is related to the flux~${\bar \gamma}$ as follows:
\begin{equation}
{\bar \gamma}=\oint\limits_{\cal P} A_{\mu}\, dx^{\mu}
=\int\limits_{0}^{2\pi} B\; d\phi = (2\pi B),  
\end{equation}
where~${\cal P}$ represents a closed path that encircles 
the string once.

We shall choose to work here in the gauge wherein the magnetic 
flux is represented by the vector potential~(\ref{eqn:vp}).
The advantage of working in this gauge is that the two-point 
function of a quantum field around the flux carrying string
can be evaluated directly without any recourse to imposing 
additional boundary conditions on the field.
In such a gauge, the symmetric two-point function in the 
vacuum state for the case of a massless, complex scalar 
field~${\hat \Phi}$ around the cosmic string is given by 
(for details, see App.~\ref{app:tptfn})
\begin{eqnarray}
G_{\rm v}^{(1)}({\tilde x}_{1}, {\tilde x}_{2})
&=&\left(4\pi^2\alpha \rho_{1} \rho_{2}\right)^{-1}
\left({\rm sinh}(\gamma \eta/\alpha)\; 
e^{-\left[i\left(\theta_{1}-\theta_{2}\right)/\alpha\right]}
+{\rm sinh}\left[(1-\gamma)\eta/\alpha\right]\right)\nonumber\\
& &\qquad\qquad\qquad\qquad\qquad\times\;
\biggl({\rm sinh}\eta\, \left[{\rm cosh}(\eta/\alpha) 
-\cos\left[\left(\theta_{1}-\theta_{2}\right)/\alpha\right]
\right]\biggl)^{-1},
\label{eqn:tptfnv}
\end{eqnarray}
where $\theta=(\alpha \phi)$, $\gamma=({\bar \gamma}/\gamma_{0})$, 
$\gamma_{0}$ being the flux quantum $(2\pi/e)$ and $\eta$ 
is given by Eq.~(\ref{eqn:eta}).
It should be pointed out here that it is only the fractional part 
of $\gamma$ that lies in the interval $0 < \gamma < 1$ that leads 
to non-trivial effects~\cite{se85,go90,bo91,pa91,sb98,si99} and the 
special case of $\gamma=(1/2)$ corresponds to that of a twisted 
scalar field~\cite{sm90,ro93}.

Now, consider the case wherein ${\bar \gamma}=0$ and $\alpha=(1/2)$.
In such a case, the two-point function $G_{\rm v}^{(1)}({\tilde x}_{1}, 
{\tilde x}_{2})$ above reduces to~\cite{ds88}
\begin{equation}
G_{\rm v}^{(1)}({\tilde x}_{1}, {\tilde x}_{2})
= \left(\pi^2 \rho_{1} \rho_{2}\right)^{-1} 
\left({\rm cosh}\eta\right)\, 
\biggl({\rm cosh}(2\eta) 
-\cos\left[2\left(\theta_{1}-\theta_{2}\right)\right]\biggl)^{-1}.
\end{equation}
This can be rewritten as [cf.~Eq.~(\ref{eqn:eta})]
\begin{eqnarray}
& &
\lefteqn{G_{\rm v}^{(1)}({\tilde x}_{1}, {\tilde x}_{2})}\nonumber\\
& &\qquad=\left(2\pi^2\right)^{-1} 
\biggl(\left[\left(\rho_{1}^2+\rho_{2}^2\right)
-\left(t_{1}-t_{2}\right)^2
+\left(z_{1}-z_{2}\right)^2 - 2 \rho_{1}\rho_{2}\,
\cos\left(\theta_{1}-\theta_{2}\right)\right]^{-1}\nonumber\\
& &\qquad\qquad\qquad\qquad\quad
+\left[\left(\rho_{1}^2+\rho_{2}^2\right)
-\left(t_{1}-t_{2}\right)^2
+\left(z_{1}-z_{2}\right)^2 + 2 \rho_{1}\rho_{2}\,
\cos\left(\theta_{1}-\theta_{2}\right)\right]^{-1}\biggl).
\end{eqnarray}
Notice that the first term in this expression corresponds to 
the two-point function in the Minkowski vacuum.
The symmetric two-point function at a finite temperature~$\beta^{-1}$ 
can now be expressed as the infinite image sum of the above vacuum 
two-point function (see, for e.g., Ref.~\cite{bd82}, Sec.~2.7).
It is given by~\cite{ds88}
\begin{eqnarray}
& &\!\!
\lefteqn{G_{\beta}^{(1)}({\tilde x}_{1}, {\tilde x}_{2})}\nonumber\\
& &\;\,=\left(2\pi^2\right)^{-1} 
\sum\limits_{n=-\infty}^{\infty}
\biggl(\left[\left(\rho_{1}^2+\rho_{2}^2\right)
-\left(t_{1}-t_{2}+in\beta\right)^2
+\left(z_{1}-z_{2}\right)^2 - 2 \rho_{1}\rho_{2}\,
\cos\left(\theta_{1}-\theta_{2}\right)\right]^{-1}\nonumber\\
& &\qquad\qquad\qquad
+\left[\left(\rho_{1}^2+\rho_{2}^2\right)
-\left(t_{1}-t_{2}+in\beta\right)^2
+\left(z_{1}-z_{2}\right)^2 + 2 \rho_{1}\rho_{2}\,
\cos\left(\theta_{1}-\theta_{2}\right)\right]^{-1}\biggl),
\label{eqn:tptfnft}
\end{eqnarray}
where we have used the subscript~$\beta$ to denote the fact that
the two-point function has been evaluated at a finite temperature.
Evidently, the first term corresponds to the thermal two-point
function in Minkowski spacetime. 

The mean values of the four-current density and the stress-energy 
tensor in the vacuum state and at a finite temperature can be 
expressed as
\begin{eqnarray}
\left\langle {\hat J}_{\mu} \right\rangle
&=&\lim_{2\to 1}\; {\cal J}_{\mu}^{(1,2)}\;
G^{(1)}({\tilde x}_{1}, {\tilde x}_{2}),\label{eqn:frcrrnt}\\
\left\langle {\hat T}_{\mu\nu} \right\rangle
&=&\lim_{2\to 1}\; {\cal T}_{\mu\nu}^{(1,2)}\;
G^{(1)}({\tilde x}_{1}, {\tilde x}_{2}),\label{eqn:set}
\end{eqnarray}
where $G^{(1)}({\tilde x}_{1}, {\tilde x}_{2})$ refers to the 
corresponding vacuum or finite temperature two-point function.
The differential operator ${\cal J}_{\mu}^{(1,2)}$ appearing 
in the expression above is defined as
\begin{equation}
{\cal J}_{\mu}^{(1,2)}
\equiv\left(ie/2\right)\; \left(D_{\mu}^1-D_{\mu}^{2*}\right),
\end{equation}
where the derivative $D_{\mu}^1$ [cf.~Eq.~(\ref{eqn:Dmu})] acts 
on the point ${\tilde x}_{1}$ and $D_{\mu}^{2*}$ on ${\tilde x}_{2}$.
For the case of a minimally coupled, massless, complex scalar field 
we are considering here, the operator~${\cal T}_{\mu\nu}^{(1,2)}$ is 
given by
\begin{equation}
{\cal T}_{\mu\nu}^{(1,2)}
\equiv \left(\frac{1}{2}\right)
\biggl(\left(D_{\mu}^1D_{\nu}^{2*}
+D_{\nu}^1D_{\mu}^{2*}\right) 
-\left(\frac{g_{\mu\nu}}{2}\right)\;
\left[g^{\kappa\lambda}\,
\left(D_{\kappa}^{1}D_{\lambda}^{2*}
+D_{\lambda}^1D_{\kappa}^{2*}\right)\right]\biggl).
\end{equation}

In the vacuum state, the mean values around the flux 
carrying string can be easily obtained using the 
two-point function~(\ref{eqn:tptfnv}).
We find that the mean value of the four-current density 
is given by [note that ${\tilde x}\equiv (t,\rho,\phi,z)$]
\begin{equation}
\left\langle {\hat J}_{\mu} \right\rangle_{\rm v}
=-\left(e\, {\cal C}/12\pi^2 \rho^2\right)\, 
\left[0,0,1,0\right],
\end{equation}
where 
\begin{equation}
{\cal C}=\left[\gamma \left(1-\gamma\right) 
\left(1-2\gamma\right)/{\alpha}^2\right].\label{eqn:calC}
\end{equation}
It should be pointed out that, in addition to the trivial 
cases of $\gamma=0$ and~$1$, the current vanishes for 
$\gamma=(1/2)$ as well.
It is also interesting to note that the conical singularity 
of the cosmic string amplifies the current by a factor of 
$\alpha^{-2}$ (compare with Refs.~\cite{se85,go90,bo91,sb98,si99} 
wherein the case of $\alpha=1$ is considered). 
The mean value of the stress-energy tensor around the string
is given by~\cite{fs87,gl94}
\begin{equation}
\left\langle {\hat T}^{\mu}_{\nu} \right\rangle_{\rm v}
=\left(720 \pi^2 \rho^4\right)^{-1}
\biggl({\cal A}\; 
{\rm diag}\, \left[1,1,-3,1\right] 
-{\cal B}\; 
{\rm diag}\, \left[1,-(1/2),(3/2),1\right]\biggl),
\end{equation}
where 
\begin{eqnarray}
{\cal A}&=& \biggl(\left[1-\left(1/\alpha^{4}\right)\right]
+ 30\, \left[\gamma^2\left(1-\gamma\right)^{2}/
\alpha^4\right]\biggl),\label{eqn:calA}\\
{\cal B}&=& 20\, \biggl(\left[1-\left(1/\alpha^{2}\right)\right] 
+ 6\, \left[\gamma \left(1-\gamma\right)/
\alpha^{2}\right]\biggl).\label{eqn:calB}
\end{eqnarray}  

Let us now consider the mean values at a finite temperature
around the string for the special case wherein ${\bar \gamma}=0$ 
and $\alpha=(1/2)$.
When the flux is absent, then, obviously, no current 
will be induced. 
The mean value of the stress-energy tensor can be 
obtained using the finite temperature two-point 
function~(\ref{eqn:tptfnft}).
We find that the sums involved can be expressed in terms 
of elementary functions (cf.~Ref.~\cite{pbm86}, Vol.~1,
pp.~687--688) and the stress-energy density is given 
by~\cite{ds88}
\begin{equation}
\left\langle {\hat T}^{\mu}_{\nu} \right\rangle_{\beta}
=\biggl({\cal A}_{\beta}\; {\rm diag}\left[1,-(1/3),-(1/3),-(1/3)\right] 
+{\cal B}_{\beta}\; {\rm diag} \left[1,-1,2,0\right]
+{\cal C}_{\beta}\; {\rm diag}\left[0,0,1,1\right]\biggl),
\label{eqn:Tmunubeta}
\end{equation}
where
\begin{eqnarray}
{\cal A}_{\beta}&=&\left(\pi^2/ 15 \beta^4\right),\label{eqn:calAbeta}\\
{\cal B}_{\beta}&=&\left(16\pi \rho^{3} \beta\right)^{-1}
{\rm coth}\left(2\pi\rho/\beta\right)
+ \left(8 \rho^{2} \beta^{2}\right)^{-1}
{\rm cosech}^{2}\left(2\pi\rho/\beta\right),\label{eqn:calBbeta}\\
{\cal C}_{\beta}&=& \left(\pi/2\rho\beta^3\right)
{\rm coth}\left(2\pi\rho/\beta\right)\; 
{\rm cosech}^{2}\left(2\pi\rho/\beta\right).\label{eqn:calCbeta}
\end{eqnarray}
Note that the first term involving ${\cal A}_{\beta}$ in the
expression~(\ref{eqn:Tmunubeta}) above corresponds to the 
thermal stress-energy density in flat spacetime.
It should be mentioned here that we have subtracted the contribution 
due to the Minkowski vacuum in order to obtain these finite expressions 
for the mean values.

\section{Four-point functions and fluctuations}\label{sec:flctns}

In the coincidence limit, the correlation functions of the 
four-current and the stress-energy densities are given by
\begin{eqnarray}
2\, \left\langle {\hat J}_{\mu}{\hat J}_{\nu}\right\rangle
&\equiv& \left\langle \left({\hat J}_{\mu}{\hat J}_{\nu}
+ {\hat J}_{\nu} {\hat J}_{\mu}\right)\right\rangle
= \lim_{4\to 3\to 2 \to 1}\;
{\cal J}_{\mu}^{(1,2)}\; {\cal J}_{\nu}^{(3,4)}\;
{\cal G}\left({\tilde x}_{1}, {\tilde x}_{2},
{\tilde x}_{3},{\tilde x}_{4}\right),\label{eqn:frcrrntflctns}\\
2\, \left\langle {\hat T}_{\mu\nu}{\hat T}_{\kappa\lambda}\right\rangle
&\equiv& \left\langle \left({\hat T}_{\mu\nu} {\hat T}_{\kappa\lambda}
+{\hat T}_{\kappa\lambda}{\hat T}_{\mu\nu}\right)\right\rangle
= \lim_{4\to 3\to 2 \to 1}\;
{\cal T}_{\mu\nu}^{(1,2)}\;{\cal T}_{\kappa\lambda}^{(3,4)}\;
{\cal G}\left({\tilde x}_{1}, {\tilde x}_{2},
{\tilde x}_{3},{\tilde x}_{4}\right),\label{eqn:stflctns}
\end{eqnarray}
where ${\cal G}\left({\tilde x}_{1}, {\tilde x}_{2},
{\tilde x}_{3},{\tilde x}_{4}\right)$ is the four-point 
function defined as
\begin{eqnarray}
& &{\cal G}\left({\tilde x}_{1}, {\tilde x}_{2},
{\tilde x}_{3},{\tilde x}_{4}\right)\nonumber\\
& &\qquad\quad
=\;\biggl\langle \left({\hat \Phi}({\tilde x}_{1})\, 
{\hat \Phi}^{\dag}({\tilde x}_{2}) 
+ {\hat \Phi}^{\dag}({\tilde x}_{2})\,  
{\hat \Phi}({\tilde x}_{1})\right)
\left({\hat \Phi}({\tilde x}_{3}) 
{\hat \Phi}^{\dag}({\tilde x}_{4})
+ {\hat \Phi}^{\dag}({\tilde x}_{4})
{\hat \Phi}({\tilde x}_{3}) \right)\nonumber\\
& &\qquad\qquad\qquad\qquad
+\, \left({\hat \Phi}({\tilde x}_{3})\,  
{\hat \Phi}^{\dag}({\tilde x}_{4}) 
+ {\hat \Phi}^{\dag}({\tilde x}_{4})\,  
{\hat \Phi}({\tilde x}_{3}) \right)
\left({\hat \Phi}({\tilde x}_{1})\,
{\hat \Phi}^{\dag}({\tilde x}_{2})
+ {\hat \Phi}^{\dag} ({\tilde x}_{2})\,
{\hat \Phi}({\tilde x}_{1})\right)\biggl\rangle.
\end{eqnarray}
At both zero and at a finite temperature $\beta^{-1}$, using 
Wick's theorem, we can write
\begin{eqnarray}
& &\biggl\langle \left({\hat \Phi}({\tilde x}_{1})\, 
{\hat \Phi}^{\dag}({\tilde x}_{2}) 
+ {\hat \Phi}^{\dag}({\tilde x}_{2})\,  
{\hat \Phi}({\tilde x}_{1})\right)
\left({\hat \Phi}({\tilde x}_{3})\, 
{\hat \Phi}^{\dag}({\tilde x}_{4})
+ {\hat \Phi}^{\dag}({\tilde x}_{4})\,
{\hat \Phi}({\tilde x}_{3}) \right)\biggl\rangle\nonumber\\
& &\qquad\quad
=\,\left\langle {\hat \Phi}({\tilde x}_{1})\,
{\hat \Phi}^{\dag}({\tilde x}_{2})
+ {\hat \Phi}^{\dag}({\tilde x}_{2})\, 
{\hat \Phi}({\tilde x}_{1})\right\rangle
\left\langle\left({\hat \Phi}({\tilde x}_{3}) 
{\hat \Phi}^{\dag}({\tilde x}_{4})
+ {\hat \Phi}^{\dag}({\tilde x}_{4}) 
{\hat \Phi}({\tilde x}_{3})\right)\right\rangle\nonumber\\
& &\qquad\qquad\qquad\qquad\qquad\qquad\qquad\qquad
\qquad\qquad\quad
+\; 4\, \left\langle {\hat \Phi}({\tilde x}_{1})\,
{\hat \Phi}^{\dag}({\tilde x}_{4})\right\rangle\,
\left\langle {\hat \Phi}^{\dag}({\tilde x}_{2})\,
{\hat \Phi}({\tilde x}_{3})\right\rangle.\label{eqn:wthm}
\end{eqnarray}
(For a discussion on Wick's theorem in the vacuum state, 
see Ref.~\cite{iz80} and, for a discussion at a finite 
temperature, see Ref.~\cite{lo73}.)
It ought to be emphasized here that the relation~(\ref{eqn:wthm})
above is valid {\it only}\/ when the expectation values are 
evaluated in the vacuum state or at a finite temperature.
This relation will not be valid, for instance, if the quantum
field is assumed to be in a $n$-particle state or, for that 
matter, in a generalized squeezed state. 
Therefore, in the vacuum state or at a finite temperature, the
expressions~(\ref{eqn:frcrrntflctns}) and~(\ref{eqn:stflctns}) 
reduce to
\begin{eqnarray}
& &\left(\left\langle {\hat J}_{\mu}
{\hat J}_{\nu}\right\rangle
- \left\langle {\hat J}_{\mu}\right\rangle
\left\langle {\hat J}_{\nu}\right\rangle\right)\nonumber\\
& &\qquad\quad
= \lim_{4\to 3\to 2 \to 1}\;2\;
{\cal J}_{\mu}^{(1,2)}\; {\cal J}_{\nu}^{(3,4)}\;
\left[G^{+}\left({\tilde x}_{1},{\tilde x}_{4}\right)\;
G^{-}\left({\tilde x}_{3},{\tilde x}_{2}\right)
+G^{-}\left({\tilde x}_{1},{\tilde x}_{4}\right)\;
G^{+}\left({\tilde x}_{3},{\tilde x}_{2}\right)\right],\\
& &\left(\left\langle {\hat T}_{\mu\nu}
{\hat T}_{\kappa\lambda}\right\rangle
- \left\langle {\hat T}_{\mu\nu}\right\rangle
\left\langle {\hat T}_{\kappa\lambda}\right\rangle\right)\nonumber\\
& &\qquad\quad
= \lim_{4\to 3\to 2 \to 1}\;2\; 
{\cal T}_{\mu\nu}^{(1,2)}\; {\cal T}_{\kappa\lambda}^{(3,4)}\;
\left[G^{+}\left({\tilde x}_{1},{\tilde x}_{4}\right)\;
G^{-}\left({\tilde x}_{3},{\tilde x}_{2}\right)
+G^{-}\left({\tilde x}_{1},{\tilde x}_{4}\right)\;
G^{+}\left({\tilde x}_{3},{\tilde x}_{2}\right)\right],
\end{eqnarray}
where $G^{\pm}\left({\tilde x}_{a},{\tilde x}_{b}\right)$ 
refer to the corresponding vacuum or finite temperature 
Wightman functions [cf.~Eq.~(\ref{eqn:wfns})].

Let us now evaluate the variance (i.e. the mean-squared deviations) 
in the current (${\hat J}_{\phi}$) and the energy (${\hat T}_{tt}$) 
densities.
For convenience, we shall hereafter denote ${\hat J}_{\phi}$ 
as ${\hat j}$ and ${\hat T}_{tt}$ as~${\hat \varepsilon}$.
On using the two-point function~(\ref{eqn:tptfnv}) to 
represent the Wightman functions $G_{\rm v}^{\pm}\left({\tilde x}_{a},
{\tilde x}_{b}\right)$ (which can be done with a suitable 
introduction of a factor of $(i\epsilon)$, where 
$\epsilon\to 0^{+}$), it can be shown that the variance in the 
current and energy densities in the vacuum state are given by
\begin{equation}
\left(\left\langle {\hat j}^2 \right\rangle_{\rm v}
- \left\langle {\hat j}\right\rangle_{\rm v}^2\right)
=\left(e/24 \pi^2 \rho^2\right)^{2}\;
\left[2{\cal C}^2-\left(\alpha^2{\cal A}{\cal B}/400\right)\right]
\end{equation}
and 
\begin{equation}
\left(\left\langle {\hat \varepsilon}^2\right\rangle_{\rm v}
-\left\langle {\hat \varepsilon}\right\rangle_{\rm v}^2\right)
=\left(1440\pi^2\rho^4\right)^{-2}\;
\left[12{\cal A}^2+ \left(9{\cal B}^2/2\right)
+6{\cal A}{\cal B}+ \left(7200\, {\cal C}^2/\alpha^2\right)\right],
\end{equation}
where ${\cal C}$ is given by Eq.~(\ref{eqn:calC}) and ${\cal A}$ and 
${\cal B}$ are given by Eqs.~(\ref{eqn:calA}) and~(\ref{eqn:calB}).
For the case ${\bar \gamma}=0$ and $\alpha=(1/2)$, the mean-squared
deviations in the energy density at a finite temperature can similarly 
be evaluated using the two-point function~(\ref{eqn:tptfnft}). 
We find that 
\begin{equation}
\left(\left\langle {\hat \varepsilon}^2\right\rangle_{\beta}
-\left\langle {\hat \varepsilon}\right\rangle_{\beta}^2\right)
=\left[\left({\cal A}_{\beta}^2/3\right) 
+\left(3{\cal B}_{\beta}^2/2\right)
+\left({\cal C}_{\beta}^2/2\right)
+\left({\cal A}_{\beta} {\cal B}_{\beta}/3\right)
+{\cal B}_{\beta} {\cal C}_{\beta}
-\left({\cal A}_{\beta} {\cal C}_{\beta}/3\right)\right],
\end{equation}
where ${\cal A}_{\beta}$, ${\cal B}_{\beta}$ and ${\cal C}_{\beta}$
are given by Eqs.~(\ref{eqn:calAbeta}), (\ref{eqn:calBbeta}) and
(\ref{eqn:calCbeta}), respectively.

At this stage of our discussion, it is important that we comment 
on the procedure we have adopted here to regularize the four-point 
functions and their derivatives in the limit when all the four 
points coincide. 
Towards the end of the last section, we had mentioned that, in 
order to obtain divergence free expressions for the mean values, 
we had regularized the quantities involving the two-point functions 
and their derivatives (in the coincidence limit) by subtracting 
the corresponding contribution due to the Minkowski vacuum. 
Since, using Wick's theorem, we can express the four-point 
functions (both in the vacuum state and at a finite temperature) 
in terms of the two-point functions, the divergences in the 
four-point functions and their derivatives (when all the four 
points coincide) essentially involve the divergences of the 
two-point functions and their derivatives.     
Therefore, evidently, if we use the regularized two-point 
functions to evaluate the corresponding four-point functions, 
then the resulting four-point functions will be free of the 
divergences in the coincidence 
limit~\cite{kf93,ph97,wf99,ph00b,hp00,ph00a}.
Indeed, this is the regularization procedure we have adopted to 
obtain divergence free expressions for the mean-squared deviations. 

\section{Magnitude of fluctuations}\label{sec:mag}

A useful measure of the magnitude of fluctuations in a 
stochastic variable is the dimensionless ratio of the 
variance to the mean-squared value of the variable.
For a fluctuating quantum variable that is represented by the 
operator~${\hat {\cal O}}$, such a dimensionless quantity can 
be defined as~\cite{kf93,ph97,wf99,ba91a,hp00,ph00a}
\begin{equation}
\Delta_{\cal O}
=\left(\frac{\left\langle {\hat {\cal O}}^2\right\rangle 
- \left\langle {\hat {\cal O}} \right\rangle^2}
{\left\langle {\hat {\cal O}}^2 \right\rangle}\right),
\end{equation}
where the expectation values are evaluated in a given state.
The fluctuations in the quantity~${\cal O}$ can be considered 
to be large if $\Delta_{\cal O}\simeq 1$ and the fluctuations 
can be said to be small if $\Delta_{\cal O} \ll 1$.

From the results we have obtained in the last two sections, it is 
easy to show that, in the vacuum state, the quantity $\Delta$ 
corresponding to the current and energy densities are given by   
\begin{equation}
\Delta_{j}
=\left[2\, {\cal C}^2
-\left(\alpha^2{\cal A}{\cal B}/400\right)\right]
\left[6\, {\cal C}^2
-\left(\alpha^2{\cal A}{\cal B}/400\right)\right]^{-1}
\end{equation}
and
\begin{eqnarray}
& &\Delta_{\varepsilon}
=\left[24{\cal A}^2+ 9{\cal B}^2 + 12{\cal A}{\cal B}
+\left(14400\, {\cal C}^2/\alpha^2\right)\right]\nonumber\\
& &\qquad\qquad\qquad\qquad\qquad\times\;
\left[32{\cal A}^2+ 17{\cal B}^2 -4{\cal A}{\cal B}
+\left(14400\, {\cal C}^2/\alpha^2\right)\right]^{-1}.
\end{eqnarray}
At a finite temperature~$\beta^{-1}$, the relative magnitude
of the variance in the energy density with respect to the 
mean-squared value [for the special case of ${\bar \gamma}=0$ 
and $\alpha=(1/2)$] is given by
\begin{eqnarray}
\Delta_{\varepsilon}^{\beta}
&=&\left[\left({\cal A}_{\beta}^2/3\right) 
+\left(3{\cal B}_{\beta}^2/2\right)
+\left({\cal C}_{\beta}^2/2\right)
+\left({\cal A}_{\beta} {\cal B}_{\beta}/3\right)
+{\cal B}_{\beta} {\cal C}_{\beta}
-\left({\cal A}_{\beta} {\cal C}_{\beta}/3\right)\right]\nonumber\\
& &\qquad\quad\times\;
\left[\left(4{\cal A}_{\beta}^2/3\right) 
+\left(5{\cal B}_{\beta}^2/2\right)
+\left({\cal C}_{\beta}^2/2\right)
+\left(7{\cal A}_{\beta} {\cal B}_{\beta}/3\right)
+{\cal B}_{\beta} {\cal C}_{\beta}
-\left({\cal A}_{\beta} {\cal C}_{\beta}/3\right)\right]^{-1}.
\end{eqnarray}

\begin{figure}[!htb]
\epsfysize=5.5 true in
\hskip 80 true pt
\epsfbox{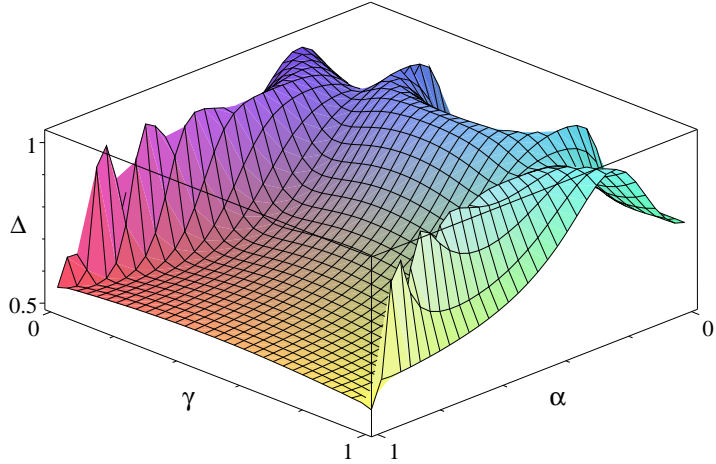}
\vskip -180 true pt
\caption{$\Delta_{\varepsilon}$ vs. $(\alpha,\gamma)$} 
\label{fig:deltae}
\end{figure}

\begin{figure}[!htb]
\epsfysize=5.5 true in
\hskip 80 true pt
\epsfbox{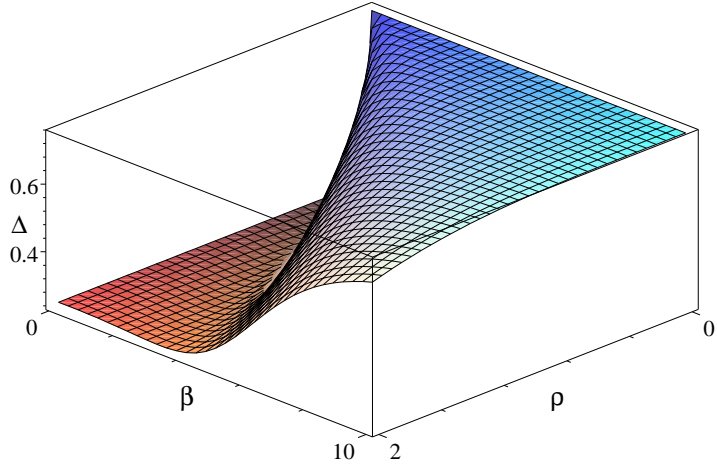}
\vskip -180 true pt
\caption{$\Delta_{\varepsilon}^{\beta}$ vs. $(\rho,\beta)$}
\label{fig:deltaeb}
\end{figure}

In Fig.~\ref{fig:deltae}, we have plotted $\Delta_{\varepsilon}$ 
for the entire range of the variables $\alpha$ and $\gamma$.
And, in Fig.~\ref{fig:deltaeb}, we have plotted 
$\Delta_{\varepsilon}^{\beta}$ for a sufficiently 
wide range of the variables $\rho$ and $\beta$. 
It is clear from these two figures that both $\Delta_{\varepsilon}$ 
and $\Delta_{\varepsilon}^{\beta}$ are always of order unity which 
suggests that the fluctuations in the energy-density around cosmic 
strings can be considered to be large.

\begin{figure}[!htb]
\epsfysize=5.5 true in
\hskip 80 true pt
\epsfbox{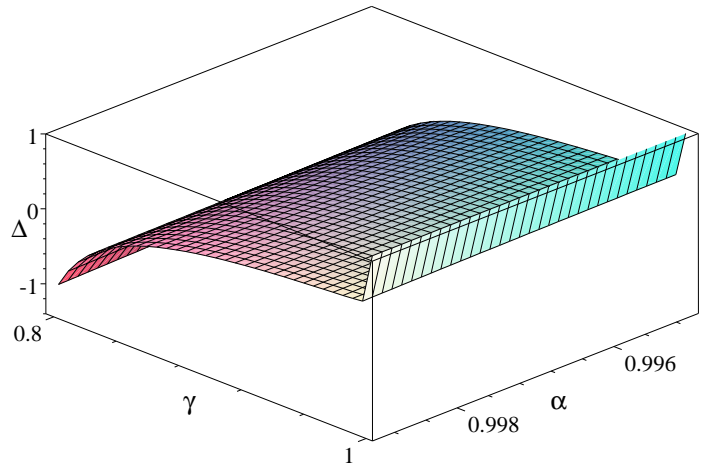}
\vskip -180 true pt
\caption{$\Delta_{j}$ vs. $(\alpha, \gamma)$}
\label{fig:deltaj}
\end{figure}

In the cases of $\Delta_{\varepsilon}$ and 
$\Delta_{\varepsilon}^{\beta}$, both the 
numerator and the denominator remain positive 
definite and also prove to be of the same order 
of magnitude for the entire range of the 
variables $(\alpha,\gamma)$ and $(\rho,\beta)$.
As a result, $\Delta_{\varepsilon}$ and 
$\Delta_{\varepsilon}^{\beta}$ turn out 
to be of order unity.  
In contrast, there exist values of $\alpha$ and 
$\gamma$ for which the numerator and the denominator
of $\Delta_{j}$ vanish.
In Fig.~\ref{fig:deltaj}, we have plotted $\Delta_{j}$ 
for a small range of the variables $\alpha$ and $\gamma$ 
in a region where the denominator is non-zero. 
Evidently, within this range, apart from those regions near
the points where it vanishes identically, the magnitude of
$\Delta_{j}$ can be said to be of order unity. 

\section{Discussion}\label{sec:dscssn}

In the absence of a length scale such as the mass of the quantum 
field, it is only natural to expect that dimensionless quantities 
such as $\Delta_{j}$ and $\Delta_{\varepsilon}$, if they do not vanish 
identically, they will turn out to be of order unity~\cite{hp00,ph00a}. 
It will be worthwhile to investigate how $\Delta_{j}$ and
$\Delta_{\varepsilon}$ behave for massive fields around
flux carrying cosmic strings. 
Also, as we had discussed earlier, we had evaluated the mean values 
and the fluctuations at the {\it same}\/ spacetime point.  
However, if one considers smeared or point-separated quantities 
rather than point-coincident ones as we have done here, then the
quantities $\Delta_{j}$ and $\Delta_{\varepsilon}$ will depend on 
the smearing or probing scale~\cite{hp00,ph00a}. 
It will be interesting to examine how these quantities behave as a
function of the smearing scale, in particular when there is another
scale present such as in the case of a massive field.  
We plan to address these issues in a future publication.

\acknowledgments{}

We would like to thank Don Page, Valeri Frolov, Andrei Zelnikov 
and Jonathan Oppenheim for discussions, Don Page and Andrei 
Zelnikov for comments on the manuscript and Jonathan Oppenheim 
for help with Maple. 
This work was supported in part by the Natural Sciences and 
Engineering Research Council of Canada. 

\appendix

\section{Evaluating the two-point function}\label{app:tptfn}

A massless, complex scalar field~$\Phi$ evolving in a 
classical electromagnetic and gravitational background 
described by the vector potential~$A^{\mu}$ and the metric 
tensor~$g_{\mu\nu}$ satisfies the differential equation
\begin{equation}
\frac{1}{\sqrt{-g}}D_{\mu}
\left(\sqrt{-g}\, g^{\mu\nu}D_{\nu}\right)\Phi=0,
\end{equation}
where the differential operator $D_{\mu}$ is defined as
\begin{equation}
D_{\mu}=\left(\partial_{\mu}+ieA_{\mu}\right).
\label{eqn:Dmu}
\end{equation}
The normalized, positive norm modes of the field 
$\Phi$ evolving in a background described by the 
line-element~(\ref{eqn:csmtrc}) and the vector 
potential~(\ref{eqn:vp}) are given by
\begin{equation}
u_{qlk_z} ({\tilde x})= 
\left(q/(2\pi)^2\, (2\omega \alpha) \right)^{1/2}\, 
e^{-i\omega t}\; e^{il\phi}\; e^{ik_{z}z}\; J_{\sigma}(q\rho),
\label{eqn:modes}
\end{equation}
where $\omega=\left(q^2+k_z^2\right)^{1/2}$ and 
$J_{\sigma}(q\rho)$ is the Bessel function of order 
$\sigma$ with $\sigma=\left(\vert(l+eB)/\alpha\vert\right)$.
The completeness relation for these modes leads to the
following conditions on the wave numbers: $0\le q < \infty$, 
$-\infty < k_z < \infty$ and $l= 0, \pm 1, \pm 2,\ldots$.

The symmetric two-point function $G^{(1)}({\tilde x}_{1}, 
{\tilde x}_{2})$ of the quantum field~${\hat \Phi}$ is 
defined as (see, for e.g., Ref.~\cite{bd82}, Sec.~2.7)
\begin{equation}
G^{(1)}({\tilde x}_{1}, {\tilde x}_{2})
=G^{+} ({\tilde x}_{1}, {\tilde x}_{2}) 
+G^{-}({\tilde x}_{1}, {\tilde x}_{2}),
\end{equation}
where $G^{+}({\tilde x}_{1}, {\tilde x}_{2})$ and 
$G^{-}({\tilde x}_{1}, {\tilde x}_{2})$ are the 
Wightman functions given by  
\begin{equation}
G^{+} ({\tilde x}_{1}, {\tilde x}_{2}) 
\equiv \left\langle {\hat \Phi}({\tilde x}_{1})\, 
{\hat \Phi}^{\dag}({\tilde x}_{2})\right\rangle,
\quad\quad
G^{-}({\tilde x}_{1}, {\tilde x}_{2}),
\equiv \left\langle {\hat \Phi}^{\dag}({\tilde x}_{2})\, 
{\hat \Phi}({\tilde x}_{1})\right\rangle\label{eqn:wfns}
\end{equation}
and the expectation values are evaluated in a given state of 
the quantum field.
In the vacuum state, the symmetric two-point function around
the flux carrying cosmic string can be expressed in terms of 
the normalized modes~(\ref{eqn:modes}) as follows:
\begin{eqnarray}
G_{\rm v}^{(1)}({\tilde x}_{1}, {\tilde x}_{2})
&=& \int\limits_{0}^{\infty}dq\, 
\sum\limits_{l=-\infty}^{\infty}\;
\int\limits_{-\infty}^{\infty}dk_{z}\,
\left(q/(2\pi)^2\, \alpha \right)\,
e^{il(\phi_{1}-\phi_{2})}\; 
e^{ik_{z}(z_{1}-z_{2})}\;
J_{\sigma}(q\rho_{1})\; J_{\sigma}(q\rho_{2})\nonumber\\
& &\qquad\qquad\qquad\qquad\qquad\quad\quad\times\;
\biggl[\left(\frac{1}{2\omega}\right)
\left(e^{-i\omega (t_{1}-t_{2})}
+e^{i\omega (t_{1}-t_{2})}\right)\biggl],\label{eqn:tpfn}
\end{eqnarray}
where we have used the subscript ${\rm v}$ to indicate the
fact that the expectation values have been evaluated in the 
vacuum state. 
The quantity in the square brackets in the above equation 
can now be written as (cf.~Ref.~\cite{gr80}, p.~307)
\begin{equation}
\left(\frac{1}{2\omega}\right)
\left(e^{-i\omega (t_{1}-t_{2})}+e^{i\omega (t_{1}-t_{2})}\right)
=\int\limits_{0}^{\infty} \frac{ds}{\sqrt{-(\pi is)}}\;
e^{-i\omega^2 s}\; e^{-\left[i(t_{1}-t_{2})^2/4s\right]}.
\end{equation}
On substituting this expression in Eq.~(\ref{eqn:tpfn}) and
integrating over $k_{z}$, we find that
\begin{eqnarray}
G_{\rm v}^{(1)}({\tilde x}_{1}, {\tilde x}_{2})
&=& \sum\limits_{l=-\infty}^{\infty}
e^{il(\phi_{1}-\phi_{2})}
\int\limits_{0}^{\infty} 
\frac{ds}{\left(4\pi^2 \alpha s\right)}\;
e^{-\left(i\left[(t_{1}-t_{2})^2 
-(z_{1}-z_{2})^2\right]/4s\right)}\nonumber\\
& &\qquad\qquad\qquad\qquad\qquad\qquad\quad\quad\times\;
\int\limits_{0}^{\infty}dq\;q\;  e^{-iq^2 s}\;
J_{\sigma}(q\rho_{1})\; J_{\sigma}(q\rho_{2}).
\end{eqnarray}
On carrying out the integrals over~$q$ and~$s$ (see, for e.g., 
Ref.~\cite{pbm86}, Vol.~2, p.~223 and p.~303), we obtain that
\begin{eqnarray}
G_{\rm v}^{(1)}({\tilde x}_{1}, {\tilde x}_{2})
&=&\sum\limits_{l=-\infty}^{\infty}\; e^{il(\phi_{1}-\phi_{2})}\,
\int\limits_{0}^{\infty} 
\frac{ds}{\left(8\pi^2 i \alpha s^2\right)}\,
e^{i\delta/s}\; I_{\sigma}(\rho_{1}\rho_{2}/2is)\nonumber\\
&=& \left(4\pi^2 \alpha \rho_{1} \rho_{2}\; 
{\rm sinh \eta}\right)^{-1}
\sum\limits_{l=-\infty}^{\infty}\; e^{il(\phi_{1}-\phi_{2})}
e^{-\sigma \eta},
\end{eqnarray}
where $I_{\sigma}(\rho_{1}\rho_{2}/2is)$ is the modified Bessel
function of order $\sigma$ and 
\begin{equation}
{\rm cosh}\eta = \left(2\delta/\rho_{1}\rho_{2}\right)
\quad\;{\rm with}\quad\;
(4\delta) = \left[\left(\rho_{1}^2+\rho_{2}^2\right)
-\left(t_{1}-t_{2}\right)^2
+\left(z_{1}-z_{2}\right)^2\right].\label{eqn:eta} 
\end{equation}
The sum over $l$ can now be evaluated to yield 
\begin{eqnarray}
G_{\rm v}^{(1)}({\tilde x}_{1}, {\tilde x}_{2})
&=&\left(4\pi^2\alpha \rho_{1} \rho_{2}\right)^{-1}
\left({\rm sinh}(eB \eta/\alpha)\; 
e^{-i\left(\phi_{1}-\phi_{2}\right)}
+{\rm sinh}\left[(1-eB)\eta/\alpha\right]\right)\nonumber\\
& &\qquad\qquad\qquad\qquad\qquad\quad\times\;
\biggl({\rm sinh}\eta\, \left[{\rm cosh}(\eta/\alpha) 
-\cos\left(\phi_{1}-\phi_{2}\right)\right]\biggl)^{-1},
\end{eqnarray}
which is the result [viz.~Eq.~(\ref{eqn:tptfnv})] we have quoted 
in the text.
It is easy to see that the two-point functions that have been 
obtained earlier in literature (see, for e.g., Ref.~\cite{gl94},
Eq.~(4.8); also see Ref.~\cite{bo91}, Eq.~(21) for the case 
of $\alpha=1$) are related to the above two-point function 
by a suitable gauge-transforming phase factor 
(i.e.~$\exp\left[-ieB\left(\phi_{1}-\phi_{2}\right)\right]$). 

\end{document}